\documentclass[preprint,12pt]{elsarticle}

\usepackage{amsmath}
\usepackage{amssymb}
\usepackage{amsthm}
\usepackage{subfigure}
\usepackage{mdwlist}
\usepackage{hyperref}
\usepackage{color}
\usepackage{comment}

\newtheorem{theorem}{Theorem}
\newtheorem{proposition}{Proposition}
\newtheorem{lemma}[theorem]{Lemma}
\newtheorem{corollary}[theorem]{Corollary}

\theoremstyle{definition}

\theoremstyle{remark}

\journal{Operations Research Letters}

\begin{document}
\begin{frontmatter}

\title{Optimal Digital Product Maintenance with a Continuous Revenue Stream} 

\author[juf]{James Fan}
\author[chg]{Christopher Griffin}
\address[juf]{
    Smeal College of Business, 
    The Pennsylvania State University
    University Park, PA 16802,
    E-mail: jamesfan@psu.edu
}
\address[chg]{
    Mathematics Department,
    United States Naval Academy,
    Annapolis, MD 21402, 
    E-mail: griffinch@ieee.org
}

\begin{abstract}

We use a control framework to analyze the digital vendor's profit maximization problem. The vendor captures market share by focusing costly effort on post-launch product maintenance, which influences user perception of the product and drives a revenue stream associated with product use. Our theoretical results show necessary and sufficient conditions for product maintenance to decline over a product's life-cycle, thus showing conditions when Lehman's 7th law of software evolution holds. We also numerically illustrate control paths under different market conditions.
\end{abstract}

\end{frontmatter}

\section{Introduction}

Digital distribution platforms for software applications represent a large and continuously growing environment for product distribution.  This is an environment where software vendors experience unique benefits and challenges. For example, because software quality may change post-launch, firms can look to maximize market share through focus on product maintenance. A strong focus on maintenance is recognized; however, maintaining this dedication is expensive and faces decreasing returns over time.  Consequently, developers must balance current and future effort costs to maximize revenue over a product's life-cycle.  

We believe this issue is especially relevant now, as digital distribution platforms (e.g. iTunes, GooglePlay, Steam) have revolutionized the way consumers purchase digital products like music, mobile applications, and video games. These platforms also represent a significant venue of profit; for example, Apple's iTunes recorded over \$4 billion in revenue during the first quarter in 2014 alone \cite{Apple}. While digital markets represent a great opportunity for potential sales, firms looking to maximize revenue from digitally-distributed software must first understand the digital market and its particular challenges.

In this paper, we model the problem of resource allocation for maintaining a product (e.g., software) that is being delivered through a digital medium (e.g., an online store). We assume the product provides a continuous revenue stream over the course of its lifetime (e.g., a digital subscription model), and that the firm has a vested interest not only in total profit, but also in the scrap value from total end-of-life market share. In particular, this market share may determine net initial revenue from its next product release. We model this problem as an optimal control problem in which effort towards product maintenance (e.g., in-version bug fixes, minor enhancements, etc.) affects not only the revenue stream but also consumer uptake. 

Our theoretical and numerical findings demonstrate that, in the face of increasing marginal costs on production maintenance, firms should steadily lower their focus on maintenance over a product's life-cycle as long as the perceived quality of the product reaches a certain threshold.  These results are supported by established literature on software evolution; as such, our model provides an analytic underpinning for Lehman's 7th law of software evolution, which states that "the quality of an E-type system will appear to be declining unless it is rigorously maintained and adapted to operational environment changes" \cite{Lehman1997}. This law is often nicknamed the "declining quality" law in the literature.  

The major contributions of this paper are: (i) We use a parsimonious optimal control model of digital product maintenance to provide analytic underpinnings for Lehman's 7th law of software evolution under general conditions; (ii) we show that the structure of this model leads to a simplified set of necessary conditions on the optimal control and (iii) we use these simplified necessary conditions to derive analytical results and rules-of-thumb for firms. Lastly, (iv) we use these results along with the matrix Riccati equation to derive necessary and sufficient conditions for optimality in the form of a system of differential equations with boundary conditions.

\section{Related Work}
\label{sec:Lit}
Modeling the development of large software systems has been an on-going area of research since the 1970's \cite{Wolverton1974}. Models of reliability of software that include costs have also been investigated \cite{Zahedi1991} in the literature since the early 90s. Even before the advent of modern digital eco-systems, digital distribution has been a popular topic of research in software engineering and management science in the past two decades. This paper investigates the optimal policy for post-launch maintenance of a software product to maximize firm revenue; as such, our research sits at the intersection of software engineering and operations research. Therefore, in this section we highlight the two streams of literature pertinent to our paper, as well as our contribution to the literatures .

Within the software engineering literature, a lot of attention has been given to the quality of software systems over time. The seminal work by Lehman proposes eight laws of software evolution that have been the subject of rigorous empirical research \cite{Lehman1997}.  For example, Munson notes that developing metrics for software measurement is a key element in software engineering \cite{Munson1995}. The author proposes a precise definition for software fault as a means of measuring declining quality in a software system and offers empirical support for the 7th law.  Other studies also find empirical support for the law of declining quality in traditional software settings \cite{Johari2011} \cite{Drouin2013} \cite{Neamtiu2012} \cite{Yu2013} as well as the mobile applications setting \cite{Zhang2013}. To our knowledge, our paper is the first to provide analytic support for the frequently observed 7th law.

Within the operations management (OM) literature, optimal control models have been used to study software development \cite{Ji2005}, enhancement and lifetime of software systems \cite{Ji2011}, and open source development \cite{Haruvy2008}. Dynamic optimization problems are also used in cooperative advertising research (\cite{He2008,He2011}), which studies the strategic interactions between two firms in a differential game. Our paper uses control theory to study the optimal path of post-launch product maintenance to maximize firm profits. Our key finding provides analytic support for a well-studied empirical law of software evolution.

\section{Model and Analytic Results}
\label{sec:Model}
We consider a single firm selling a product on digital distribution platforms. Market share changes according to a maintenance-mediated rumor-spreading model described below. The firm's objective is to maximize a combination of market share and profit during the product's post-launch life-cycle by controlling time allotted to product maintenance (rather than, e.g., developing a new product or version).
  
Define $u_t:\mathbb{R}_+ \rightarrow [0,\infty)$ to be the single valued function of time that captures all efforts related to maintaining a quality product, with $u_t = 0$ representing no effort towards product maintenance.  Let $x_t \in [0,1]$ denote the proportion of users who have adopted the vendor's product at time $t$, i.e., the proportional market share of the firm at time $t$. For notational simplicity, we drop the subscript $t$. We first consider the more general control problem faced by the firm, then provide a specific example with linear functions for revenue stream $R(u)$ and value function of product maintenance $\pi(u$).

\subsection{General Case}

The firm's optimal control problem is:
\begin{equation}\left\{
\begin{aligned}
\max\;\; &\Psi(x(T)) + \int_0^T R(u)x - Cu^2 \;\;dt\\
s.t. \;\;& \dot{x} = \pi(u) x(1-x)\\
& x(0) = x_0 > 0, \,\, u \geq 0
\end{aligned}\right.
\label{eqn:MainProblem}
\end{equation}

We assume $x(t), u(t) \in \mathcal{L}^2[0,\infty)$ and we also require $u(t)$ to be differentiable almost everywhere. We will find a smooth expression for $\dot{u}$ in the spirit of Equation 3 of \cite{B78}. In the generic form, $R(u)$ represents the revenue stream as a function of $u$, and we assume $R(u)$ is increasing, concave, twice-differentiable, and non-negative. The units of $R(u)$ are in dollars per market-share proportion. The constant $C$ is the cost coefficient of product maintenance so that costs are quadratic in maintenance effort. In our context, this accounts for the increasing marginal cost of resources as a firm dedicates more time and manpower to a product. This is shown to be a reasonable modeling assumption in digital goods (see \cite{Lahiri2013}). The function $\Psi(x(T))$ is the salvage value of market share at the terminal time $T$; i.e., the value the firm places on end-of-life market share for the product. We assume that $\Psi$ is differentiable, concave and monotonically increasing in $x(T)$.

The state dynamics are given by a modified Bass equation \cite{Bass2,Bass1994,Bass2004,Bass2004a}, which in this context is a logistic model of rumor spreading mediated by product quality. Here, $\pi(u)$ is the value function of $u$, which denotes the utility users derive from the product relative to the next best option. In our model, $\pi(u)$ governs the sign of the equation of motion - a product only experiences market growth if users prefer it to the next best alternative. We assume $\pi(u)$ is increasing, concave, and twice-differentiable. %Lastly, we assume  $\Psi(x(T)) \equiv \rho x(T)$ so that salvage value is linear. 
To derive analytic results, we use a change of variables to simplify the model. 

From the equation of motion, we have:
\begin{equation} \nonumber
\frac{dx}{x(1-x)} = \pi(u)\;dt.
\end{equation}
Assume the formal anti-derivative of the right-hand-side to be:
\begin{displaymath}
V(t) = \int \pi(u(t))\; dt.
\end{displaymath}
We can rewrite the state variable as: 
\begin{equation}
x = \frac{1}{1+A\exp\left(-V\right)},
\label{eqn:x}
\end{equation}
where $\exp\left(-V\right) = e^{-V}$, and
\begin{displaymath}
A = e^{V_0}\frac{1-x_0}{x_0} > 0
\end{displaymath}
is a constant of integration determined by $x_0$ and $V(0) = V_0$. This allows us to rewrite $\Psi$ as a function of $V(T))$:
\begin{displaymath}
  \Psi(V(T)) = \Psi\left(\frac{1}{1+A\exp\left(-V(T)\right)}\right).
\end{displaymath}
This function is monotonic in $V(T)$ and hence pseudoconcave by our assumptions on $\Psi$. Equation \ref{eqn:MainProblem} can now be written as the modified problem:
\begin{equation} \left\{
\begin{aligned}
\max\;\; & \Psi\left(\frac{1}{1+A\exp\left(-V(T)\right)}\right) + \int_0^T \frac{R(u)}{1+A\exp\left(-V\right)} - Cu^2 \;\;dt\\
s.t. \;\;& \dot{V} = \pi(u)\\
& V(0) = V_0, \,\, u \geq 0.
\end{aligned}\right.
\label{eqn:MainModified}
\end{equation}
Before proceeding, note:
\begin{enumerate}
  \item $R(u)$ is assumed to be increasing and concave.
  \item $1/\left( 1+A\exp\left(-V\right) \right)$ is increasing.
  \item $\pi(u)$ is assumed to be increasing and concave.
  \item $-Cu^2$ is concave.
\end{enumerate}
Because $R(u)$ is increasing and $u^2$ is symmetric, we can safely ignore the control constraint $u \geq 0$. We show in Equation \ref{eqn:Control1} (below) that this assumption is justified. The Hamiltonian for the modified optimal control problem is:
\begin{equation}\nonumber
  \mathcal{H} = \frac{R(u)}{1+A\exp\left(-V\right)} - Cu^2 + \lambda\pi(u),
\end{equation}
where $\lambda$ is the co-state. The co-state dynamics must satisfy:
\begin{equation}
  \frac{d\lambda}{dt} = -\frac{\partial\mathcal{H}}{\partial V}=-\frac{A \exp\left(-V\right) R(u)}{\left(A \exp\left(-V\right)+1\right)^2} = -R(u)x(1-x),
  \label{eqn:costate}
\end{equation}
and the transversality condition requires:
\begin{multline*}
  \lambda(T) = \Psi'\left(\frac{1}{1+A\exp\left(-V(T)\right)}\right)\frac{A\exp(-V(T))}{\left(1+A\exp(-V(T)) \right)^2}=\\
  \Psi'(x(T))x(T)\left(1-x(T)\right)
\end{multline*}
because:
\begin{equation}
x(1-x) = %\frac{1}{1+A\exp\left(-V\right)}\left(1 - \frac{1}{1+A\exp\left(-V\right)}\right) = 
\frac{A\exp(-V)}{\left(1+A\exp(-V) \right)^2}. 
\label{eqn:xx}
\end{equation}
Note however, this is the co-state in terms of $V$, expressed in terms of $x$. The final time value $\lambda(T)$ is positive and the time derivative of $\lambda$ is strictly negative, since $R(u) > 0$. Therefore we have:
\begin{lemma} For all time $t \geq 0$, the co-state $\lambda$ is positive and decreasing. \hfill\qed
\label{lem:costate}
\end{lemma}

% From our assumptions, we see at once that 
% \begin{displaymath}
%   \frac{R(u)}{1+A\exp\left(-V\right)}
% \end{displaymath}
% is increasing in $u$ and $V$ and therefore quasi-concave. Furthermore, we note that the scrap value $\Psi$ is also concave and non-decreasing in $V$ and therefore quasi-concave. It follows at once that $\mathcal{H}$ and $\Psi$ are quasi-concave in both $u$ and $V$. Consequently we have the following result, which follows from \cite{SS77}:
% \begin{lemma} Any solution $u^*$ to $\mathcal{H}_u = 0$ is a maximum and furthermore, if $u^*$ satisfies $\mathcal{H}_u = 0$, then this is sufficient to ensure it is an optimal control when the state and co-state are found by solving the Euler-Lagrange equations. \hfill\qed
% \end{lemma}
The Hamiltonian is (strictly) concave in the control $u$, and thus we have:
\begin{lemma} Any solution $u^*$ to $\mathcal{H}_u = 0$ satisfies the necessary conditions:
\begin{enumerate}
\item $\mathcal{H}_u = 0$ and
\item $\mathcal{H}_{uu} < 0$, the strong Legendre-Clebsch condition
\end{enumerate}
and therefore, it maximizes the Hamiltonian at all times. \hfill\qed
\label{lem:Necessary}
\end{lemma}
It is worth noting that the two conditions in Lemma \ref{lem:Necessary} along with the fact that $V^*$ ($x^*$, resp.) and $\lambda^*$ solve the resulting Euler-Lagrange two-point boundary value problem form the complete set of necessary conditions for the optimal control problem. When we add the additional requirement that the corresponding Riccati equation is bounded on $[0,T]$, these form sufficient conditions for a weak local maximal optimal controller \cite{J70}. We discuss this sufficient condition in Section \ref{sec:Suff}.

Given the necessary conditions of the optimal control problem, we study solutions to $\mathcal{H}_u = 0$. A solution to $\mathcal{H}_u = 0$ yields the implicit equation:
\begin{multline}
  u(t)=\frac{1}{2C}\left(\frac{R'(u(t))}{A \exp\left(-V(t)\right)+1}+\lambda (t) \pi'(u(t))\right) =\\ 
  \frac{1}{2C}\left(R'(u(t))x(t) + \lambda(t)\pi'(u(t))\right)
  \label{eqn:Control1}
\end{multline}
%By an extension of the Mangaserian sufficiency theorem to quasi-concave functions \cite{SS77}, the optimal control must satisfy this relationship, and it is a sufficient condition. 
We note also that $u^*(t) \geq 0$ for all $t$, since $R'(u), \pi'(u) > 0$. Thus our assumption to ignore the constraint $u \geq 0$ is now justified.

\begin{proposition} Assume $u^*$ is a solution to the optimal control problem. Then, optimal focus on product maintenance (i.e., $u^*$) is decreasing if and only if:
\begin{equation}\nonumber
\pi(u) > \frac{\pi'(u) R(u)}{R'(u)}.
\end{equation}
\label{prop:general}
\end{proposition}
\begin{proof}
Differentiating Equation \ref{eqn:Control1} with respect to $t$ and substituting the value for $\dot{\lambda}$ in terms of $V$ (see Equation \ref{eqn:costate}) yields\footnote{Simplification performed using Mathematica\textsuperscript{TM}.}:
\begin{equation}
  \dot{u} = \frac{A e^V \left(\pi (u) R'(u)-R(u) \pi '(u)\right)}{\left(A+e^V\right)
   \left(\left(A+e^V\right) \left(2 C -\lambda (t) \pi ''(u)\right)-e^V
   R''(u)\right)}
   \label{eqn:udot1}
\end{equation}
By assumption $\pi'(u), R'(u) > 0$ and $\pi''(u), R''(u) < 0$. Therefore,
the denominator of Equation \ref{eqn:udot1} is always positive. The numerator is positive if and only if:
\begin{equation}\nonumber
  \pi (u) R'(u)-R(u) \pi '(u) > 0 \iff \pi(u) > \frac{\pi'(u) R(u)}{R'(u)}
\end{equation}
\end{proof}

Proposition \ref{prop:general} states that focus on product maintenance decreases if the valuation of the product surpasses the threshold determined by the relationship between $R(u)$, $R'(u)$, and $\pi'(u)$. That is to say, when consumers value the firm's product above the next best alternative past a certain threshold, it is in the firm's best interest to decrease focus on product maintenance. This is often observed in practice,  when one product on the market is established as the best option. Decreasing product maintenance is actually the optimal strategy for the firm as long as perceived product value is above a threshold. Proposition \ref{prop:general} also provides a general analytic result in support of Lehman's 7th law of software evolution. Because rigorous maintenance and adaption to operational environment changes are cost prohibitive, the quality of the digital good appears to be declining over time. This is not to say actual product quality must necessarily decrease over time, as effort may be drawn off to develop the next product iteration. 

In the next section, we study the case when $R(u)$ and $\pi(u)$ are linear to show that $u$ is always decreasing given this assumption. The linear case also allows us to derive a simpler form for the optimal control.

\subsection{Linear Case}\label{sec:Linear}
Assume $R(u)$ and $\pi(u)$ are linear and take the form $R(u) = \alpha u$ and $\pi(u) = \beta \cdot (Pu - \pi_0)$, where $\beta > 0$ is the coefficient of market share change. Assume $\Psi(x(T)) = \rho x(T)$, for $\rho \geq 0$, which is non-decreasing and concave, as required. A larger value for $\beta$ results in larger changes in market share. The firm's optimal control problem can be rewritten as:
\begin{equation}\left\{
\begin{aligned}
\max\;\; &\rho x(T) + \int_0^T \alpha u x - Cu^2 \;\;dt\\
s.t. \;\;& \dot{x} = \beta \cdot x(1-x)\left(Pu- \pi_0\right)\\
& x(0) = x_0, \,\, u \geq 0.
\end{aligned}\right.
\label{eqn:LinearProblem}
\end{equation}

Now, $\pi_0 > 0$ is the explicit value users derive from the next best alternative, assumed to be constant over a product's life-cycle. %We first show that $u$ is always decreasing in our linear case. 
%We find for $u^*$ by solving $\mathcal{H}_u = 0$ and obtain: 
%\begin{displaymath}
%u^*=\frac{\dot{V}+\beta\pi_0}{\beta P}. 
%\end{displaymath}
Under the change of variables:
\begin{equation}\nonumber
\frac{dV}{dt} = \beta Pu - \beta\pi_0.
\end{equation}
while Equation \ref{eqn:x} is still valid for $x$ in terms of $V$.
Using this information, we can simplify the integral in the modified problem objective function of Equation \ref{eqn:LinearProblem}:
\begin{multline*}
\int_0^T \alpha u x - Cu^2 \;\;dt = \int_0^T \frac{\alpha u}{1+A\exp\left(-V\right)} - Cu^2 \;\;dt = \\
\int_0^T \frac{\alpha}{\beta P}\frac{1}{{1+A\exp\left(-V\right)}}dV + \frac{\alpha\pi_0}{P}\frac{1}{1+A\exp\left(-V\right)}-Cu^2\;\;dt = \\
\frac{\alpha}{\beta P}\log\left(A+\exp(V)\right)\vert_0^T + \int_0^T\frac{\alpha\pi_0}{P}\frac{1}{1+A\exp\left(-V\right)}-Cu^2\;\;dt.
\end{multline*}
%This problem is linear in the state dynamics.
The scrap value for the simplified optimal control problem is:
\begin{equation}
\tilde{\Psi}(V(T)) = \frac{\rho}{1+A\exp(-V(T))} + \frac{\alpha}{\beta P}\log\left(A + \exp(V(T))\right).
\label{eqn:NewScrap}
\end{equation}
The revised Hamiltonian for this simpler problem is:
\begin{equation}\nonumber
\mathcal{H} = \frac{\alpha\pi_0}{P}\frac{1}{1+A\exp\left(-V\right)}-Cu^2 + \lambda\beta(Pu-\pi_0).
\end{equation}
The necessary conditions discussed in Lemma \ref{lem:Necessary} still hold. Thus, solving $\mathcal{H}_u = 0$ leads to the expression:
\begin{equation}
u \equiv \frac{\beta P}{2C}\lambda.
\label{eqn:ControlInLambda}
\end{equation}
%As before the Hamiltonian and scrap value are quasi-concave in $u$ and $V$. %The scrap value is:
%\begin{equation}\nonumber
%  \Psi(V(T) = \frac{\rho}{1+A\exp\left(-V(T)\right)} + \frac{\alpha}{\beta P}\log\left(1+\exp(V)\right)\vert_0^T
%\end{equation}
%We have already noted that this is sufficient to ensure $u$ is an optimal control as it satisfies the generalized convexity conditions given in \cite{SS77}. Thus the Euler-Lagrange system of differential equations will allow us to derive an optimal control:
Since the Euler-Lagrange system of differential equations are necessary conditions, we know that the state and co-state must satisfy:
\begin{equation}
\begin{aligned}
\dot{V} &= \frac{(\beta P)^2}{2C}\lambda - \beta\pi_0,\\
\dot{\lambda} &= -\frac{\alpha\pi_0}{P}\frac{A\exp(-V)}{\left(1+A\exp(-V) \right)^2}.
\end{aligned}
\label{eqn:syslinear}
\end{equation}
Using these equations, we can now find a simple system of differential equations governing $u$ and the original state variable $x$, that will necessarily be satisfied by any solution to the optimal control problem.
%Notice however, $\dot{u} = (\beta P)/(2C)\dot{\lambda}$, which is simpler than in the general case. Corollary \ref{cor:linear} describes the optimal control path in the specialized linear case.
\begin{corollary} %Assuming $R(u)$ and $\pi(u)$ are linear, $u$ is always decreasing.
The optimal state and control for Equation \ref{eqn:LinearProblem} necessarily satisfy the following two-point boundary value problem:
\begin{equation}
\left\{
\begin{aligned}
\dot{x} &= \beta x(1-x)(Pu - \pi_0)\\
\dot{u} &= -\frac{\alpha\beta\pi_0}{2C}x (1-x)\\
x(0) &= x_0\\
u(T) &= \frac{1}{2C}\left(\alpha x(T) + \beta P\rho x(T)(1-x(T))\right).
\end{aligned}\right.
\end{equation}
\label{cor:linear}
\end{corollary}
\begin{proof} The state dynamics  and initial condition are given. Substituting the result from Equation \ref{eqn:syslinear} into Equation \ref{eqn:ControlInLambda} and using Equation \ref{eqn:xx}, we obtain the expression for $\dot{u}$:
\begin{equation}
\dot{u} = -\frac{\beta P}{2C}\frac{\alpha\pi_0}{P}\frac{A\exp(-V)}{\left(1+A\exp(-V) \right)^2} = -\frac{\alpha\beta\pi_0}{2C}x(1-x).
\label{eqn:ut}
\end{equation}
Differentiating Equation \ref{eqn:NewScrap} with respect to $V$ and constructing the transversality condition yields:
\begin{multline*}
\lambda(T) = \rho\frac{A\exp(-V(T))}{\left(1+A\exp(-V(T))\right)^2} + \frac{\alpha}{\beta P}\frac{1}{1+A\exp(-V(T))} =\\ \rho x(T)(1-x(T)) + \frac{\alpha}{\beta P} x(T).
\end{multline*}
Solving for $u(T)$ using $\lambda(T)$ and Equation \ref{eqn:ControlInLambda} yields:
\begin{displaymath}
u(T) = \frac{1}{2C}\left(\beta P\rho x(T)(1-x(T)) + \alpha x(T)\right).
\end{displaymath}
This completes the proof.
%The final time condition for $u$ follows from the transversality condition: $\lambda(T) = \Psi'(x(T))$ when combined with Equation \ref{eqn:ControlInLambda}. 
\end{proof}
%We see that $(x^*,u^*)$ are the optimal state and control trajectories when they solve the system of differential in Corollary \ref{cor:linear}. 
It is clear that $\dot{u}$ should be decreasing in the linear case because $\pi'(u) = P$ and $R'(u) = \alpha$. Therefore:
\begin{equation}\nonumber
  \frac{\pi}{\pi'} = u - \frac{\pi_0} P < u = \frac{R}{R'}.
\end{equation}
This agrees with the result shown in Corollary \ref{cor:linear}, where clearly $\dot{u} \leq 0$.

%\textcolor{red}{Chris: Can you fit the below paragraph into the paper? It is the discussion that leads to Corollary 2 \& 3 and their proofs. However, we move away from talking about the differential system in terms of $\dot{u}$ and $\dot{x}$. We could alternatively remove them, but Corollary 3 leads into the numerical examples. Maybe we cut Corollary 2? Your advice is greatly appreciated here.}

The right-hand-sides of the dynamics are smooth in both $x$ and $u$. Consequently from any starting point over the time horizon $[0,T]$, there is a \textit{unique} solution curve in the $x-u$ phase space satisfying the boundary conditions. Thus, assuming $u^*$ is the form of the optimal control, this curve is the \textit{unique} solution to the problem. Furthermore, note that $\dot{u} < 0$. Thus, $u(0) > 0$ and $u(T) \geq 0$ since $x$ is constrained (by its dynamics) to remain in the interval $[0,1]$ and all parameters are assumed to be non-negative.

% \begin{corollary} If $x(0) > 0$, then $u(t) > 0$ for all $t \in [0,T]$. 
% \end{corollary}
% \begin{proof} Note that $u(T) \equiv 0$ if and only if $x(T) \equiv 0$. However $x \equiv 0$ is an equilibrium point of the state dynamics and cannot be reached in finite time unless $x(0) = 0$. Thus if $x(0) > 0$, then $u(T) > 0$. Assume $x(0) > 0$. The fact that $\dot{u} < 0$ for all time in $[0,T]$ implies that $u(0) > 0$ and $u(T) > 0$. Thus $u(t) > 0$ for all $t \in [0,T]$.
% \end{proof}

%The previous corollary proves (a posteriori) that we are justified in ignoring the constraints on the control. Additionally, 
The fact that $\dot{u} < 0$ shows that at optimality, focus on product maintenance is a monotonically decreasing function.  In the specific case where $R(u)$ and $\pi(u)$ are linear, firms should always devote the most effort towards product maintenance at the beginning of the product life-cycle and continuously scale back maintenance effort over time. 
\begin{corollary} Suppose $x$ and $u$ are the state and optimal controls. Then:
\begin{equation}
x(u) = \tilde{C} - \frac{CP}{\alpha \pi_0}u^2 + \frac{2C}{\alpha }u,
\label{eqn:xinu}
\end{equation}
where:
\begin{equation}
\tilde{C} = x(T) + \frac{CP}{\alpha \pi_0}u(T)^2 - \frac{2C}{\alpha }u(T).
\label{eqn:tildeC}
\end{equation}
That is, $x$ varies quadratically in $u$.
\label{cor:corollary4}
\end{corollary}
\begin{proof} From the system of differential equations given in Corollary \ref{cor:linear}, we can express $x$ solely as a function of $u$. Note:
\begin{equation}
\label{eqn:dxdu}
\frac{\dot{x}}{\dot{u}} = \frac{d x}{d u} = -\frac{2C}{\alpha \pi_0}\left(Pu - \pi_0\right).
\end{equation}
Integrating Equation (\ref{eqn:dxdu}) yields an expression for the state variable as a function of the control:
\begin{displaymath}
x(u) = \tilde{C} - \frac{CP}{\alpha \pi_0}u^2 + \frac{2C}{\alpha }u
\end{displaymath}
where $\tilde{C}$ is the constant of integration. We have an expression for $u(T)$:
\begin{displaymath}
u(T) = \frac{\alpha x(T) + \beta P \rho x(T) \left(1 - x(T)\right)}{2C}
\end{displaymath}
Thus we can solve for $\tilde{C}$, given in Equation \ref{eqn:tildeC}.
\end{proof}

\begin{corollary} Assume the optimal control is given by Expression \ref{eqn:ut}. Exactly one of the following holds:
\begin{enumerate}
\item $u(0) > \pi_0/P$, $u(T) \geq \pi_0/P$ and thus $x(T) \geq x(0)$ and market share increases monotonically.
\item $u(0) > \pi_0/P$, $u(T) < \pi_0/P$ and market share increases and then decreases, and the final relationship between $x(T)$ and $x(0)$ is determined by the relationship between $u(T)$ and $u(0)$. 
\item $u(0) \leq \pi_0/P$ and thus $x(T) \leq x(0)$ and market share decreases monotonically.
\end{enumerate}

\label{cor:3cases}
\end{corollary}
\begin{proof} The closed-loop expression for $x$ in $u$ (Equation \ref{eqn:xinu}) is maximized at $u = P/\pi_0$. Thus, if $u(0) > P/\pi_0$, the $x$ value increases quadratically (and monotonically) as $u$ decreases to $P/\pi_0$. At this point, $x$ begins to decrease if $u$ drops below $P/\pi_0$. (See Section \ref{sec:Numerical} for examples.)
\end{proof}

From Corollaries \ref{cor:corollary4} and \ref{cor:3cases} we can characterize the optimal path of product maintenance in three different market conditions. We illustrate these paths in Section \ref{sec:Numerical}. In the next section we show sufficiency conditions for our general model.

\subsection{Sufficient Conditions for Optimality}
\label{sec:Suff}
For this section, let $f(V,u) = \pi(u)$ and let $\mathcal{H}(V,u,\lambda)$ be the Hamiltonian for Problem \ref{eqn:LinearProblem} cast as a \textit{minimization problem}: 
\begin{displaymath}
\mathcal{H} = \frac{-R(u)}{1+A\exp(-V)} + Cu^2 + \lambda \pi(u)
\end{displaymath}
Sufficient conditions for optimal control problems are discussed extensively in \cite{BH65,Mangasarian1966,J70,P71,KS71,SS77,Z84,MO02}, with the boundedness of the (matrix) Riccati equation, which arises from an accessory minimization problem \cite{BH65,MO02}:
\begin{align*}
-\dot{S} &= \mathcal{H}_{VV} + 2f_V S - \left(\mathcal{H}_{uV} + f_uS \right)^2\mathcal{H}_{uu}^{-1}\\
S(T) &= -\Psi_{VV}\rvert_{t=T}
\end{align*}
on the interval $[0,T]$ being sufficient to ensure the identified optimal control is a (weak) local maximum assuming the conditions in Lemma \ref{lem:Necessary} are satisfied when $V$ and $\lambda$ solve the Euler-Lagrange equations. Notice we have $-\Psi_{VV}\rvert_{t=T}$ rather than $\Psi_{VV}\rvert_{t=T}$ as we have converted to a minimization problem.

In this case, we can return to the formulation in the state $x$ (rather than $V$) and develop a complete set of necessary and sufficient conditions for $u^*$ to be a (weak local) optimal control, since determining the unknown parameter $A$ in Equation \ref{eqn:x} is difficult without a closed form expression for $u(t)$. Note first that:
\begin{displaymath}
\dot{x} = \pi(u)x(1-x)
\end{displaymath}
is given and from Equation \ref{eqn:costate} we deduce that:
\begin{displaymath}
\dot{\lambda} = R(u)x(1-x)
\end{displaymath}
since the sign of $R$ is changed in the objective function. Note $\lambda$ is now negative and \textit{increasing} as a result of the sign change (in contrast to Lemma \ref{lem:costate}). From Equation \ref{eqn:udot1}, we deduce that:
\begin{displaymath}
\dot{u} = \frac{\left(R'(u)\pi(u) - \pi'(u)R(u)\right)x(1-x)}{2C - R''(u)x + \lambda\pi''(u)}.
\end{displaymath}
This yields the same criteria for the sign of $\dot{u}$ as before (which is expected), except now $\dot{u}$ is expressed in terms of the original state $x$ and the co-state $\lambda$ \textit{from the transformed problem.} Rewriting the Riccati equation in terms of $x$, $u$ and $\lambda$ yields:
\begin{displaymath}
\dot{S} = R(u)(1-2x)x(1-x) + \frac{\left(R'(u)x(1-x) - \pi'(u)S\right)^2}{2C - R''(u)x + \lambda\pi''(u)}.
\end{displaymath}
Lastly, we can compute $-\Psi_{VV}\rvert_{t=T}$ in terms of $x(T)$:
\begin{multline*}
-\Psi_{VV}\rvert_{t=T} =\\ \left(\Psi'(x(T))\left(2x(T)-1\right)-\Psi''(x(T))x(T)\left(1-x(T)\right) \right)x(T)\left(1-x(T)\right).
\end{multline*}
Finally, we have the following sufficiency result:
\begin{proposition} Suppose that $x^*$, $u^*$, $\lambda^*$ and $S^*$ solve the following system of differential equations:
\begin{equation}
\begin{aligned}
\dot{x} &= \pi(u)x(1-x),\\
\dot{\lambda} &= R(u)x(1-x),\\
\dot{u} &= \frac{\left(R'(u)\pi(u) - \pi'(u)R(u)\right)x(1-x)}{2C - R''(u)x + \lambda\pi''(u)},\\
\dot{S} &= R(u)(1-2x)x(1-x) + \frac{\left(R'(u)x(1-x) - \pi'(u)S\right)^2}{2C - R''(u)x + \lambda\pi''(u)},\\
&x(0) = x_0,\\
&\lambda(T) = -\Psi'(x(T))x(T)\left(1-x(T)\right),\\
&2Cu(T) = R'(u(T))x(T)-\lambda(T)\pi'(u(T)),\\
&S(T) = -\Psi_{VV}\rvert_{t=T},
\end{aligned}
\end{equation}
$S$ is bounded over the interval $[0,T]$ and for all time $t \in [0,T]$, and $u(t)$ satisfies the fixed point condition:
\begin{displaymath}
2C u(t) = R'(u(t))x(t)-\lambda(t)\pi'(u(t)).
\end{displaymath}
Then $(x^*,u^*)$ are a (weak local) optimal state and control for Problem \ref{prop:general}.
\label{prop:Suff}
\end{proposition}
\begin{proof}
If $(x^*,u^*,\lambda^*,S^*)$ is a solution, then the corresponding $V^*$ and $\lambda^*$ must satisfy the Euler-Lagrange equations for the modified original problem. The fact that $u^*$ satisfies the fixed point property ensures that it solves $\mathcal{H}_u = 0$, which when taken with the fact that $\mathcal{H}_{uu} > 0$ implies that the Hamiltonian is minimized. From \cite{BH65,J70}, the fact that $S$ is bounded on $[0,T]$ is sufficient to ensure that $(x^*,u^*)$ is a weak local optimal state/control solution, since the necessary conditions from Lemma \ref{lem:Necessary} are satisfied.
\end{proof}
While this is difficult to prove in closed form, it is not unreasonably to expect $S$ to be bounded in general, since the right-hand-side $\dot{S}$ is smooth in its constituent unknowns, however this must be checked to ensure optimality for certain. As in Section \ref{sec:Linear}, the conditions Proposition \ref{prop:Suff} simplify substantially in the linear case. The derivation is straight-forward and omitted for space.

\section{Numerical Example of the Linear Case}\label{sec:Numerical}
We compare the theoretical results obtained from our model in Section \ref{sec:Model} with a more complex model that does not yield a simple theoretical analysis. The objective is to show the rules of thumb obtained for the simpler model hold in the more complex case. In particular, we add a discount factor $r$ to the objective functional to more accurately reflect the time-value of money. We also consider the more complex general Bass model \cite{Bass2,Bass1994,Bass2004,Bass2004a} with an innovation coefficient $A > 0$. The resulting optimal control problem is then:
\begin{equation}
\left\{
\begin{aligned}
\max \, &\int_0^T \! e^{-rt} [\alpha u  x  - C u^2] \, \mathrm{d}t \\
s.t. \, \, &\dot{x} = \beta(A(1-x)+ x (1-x)) \left(\pi(u) - \pi_0\right) \\
&u_t >0,\, x \in [0,1],\, x(0)=x_0 \\
\end{aligned}\right.
\label{eqn:DiscountedModel}
\end{equation}
For simplicity, we set $\rho \equiv 0$ so that the firm in question is \textit{not} explicitly concerned with market share at the end of the maintenance period. The remaining parameters are given the following values:
\begin{displaymath}
T = 4;\; \alpha = 2;\; \beta = 0.5; \pi_0 = 0.5;\; P = 1;\; r = 0.05;\; C = 1;\; A = 0.05;\;
\end{displaymath}

We divide the figures into the three cases presented in Corollary \ref{cor:3cases}.  Each case represents a different starting market share; we see that the optimal control path for the focus on product maintenance heavily depends on $x_0$.  
\begin{figure}[htbp]
\centering
   \begin{subfigure}{}
     \includegraphics[width=5cm]{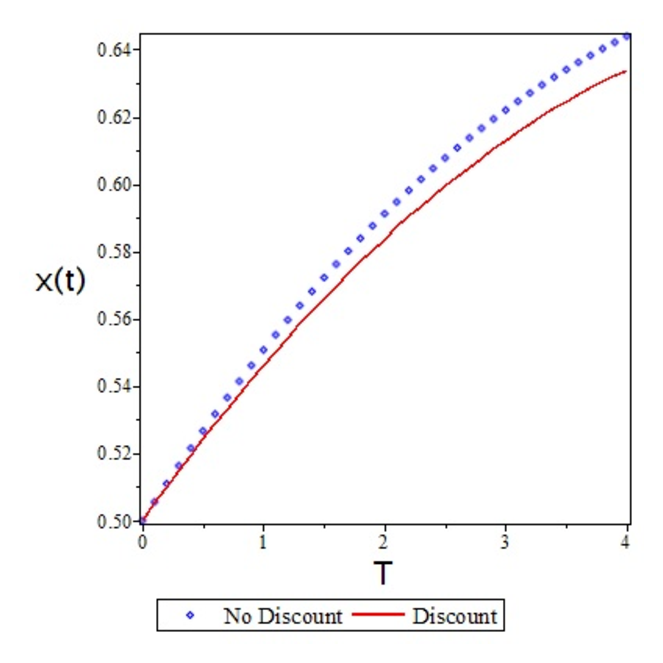}
   \end{subfigure}
\vspace{.5 cm}
   \begin {subfigure}{}
\centering
     \includegraphics[width=5cm]{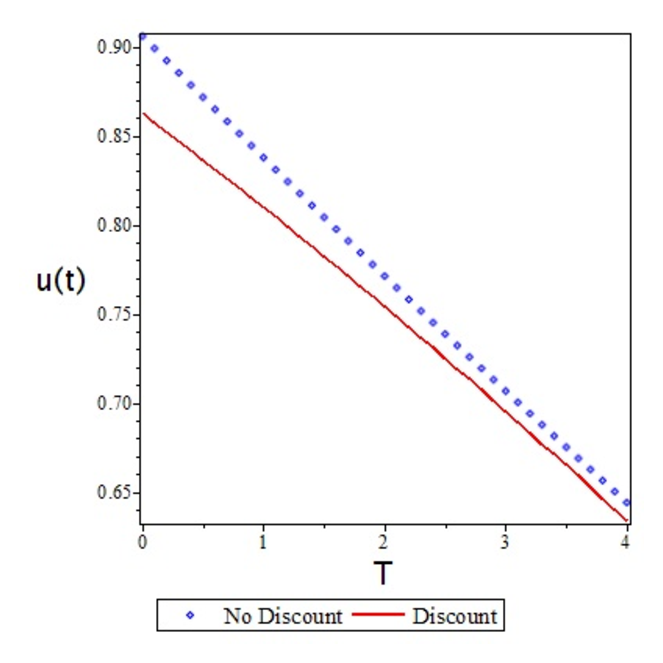}
   \end{subfigure}
\caption{Market conditions with $x_0 = 0.5$}\nonumber
\label{Fig:base05}
\end{figure}

In Figure \ref{Fig:base05}, we see the optimal control path for product maintenance and market share for the firm under Case 1 of Corollary \ref{cor:3cases}.  At this level of market share, it is profitable to maintain a relatively high level of maintenance to maximize revenue.  Market share monotonically increases for the product life-cycle, even though product maintenance is decreasing over time, consistent with Lehman's 7th law.  Case 1 represents a vendor with a large, established market share. With a strong dedication to maintenance, the firm is able to secure more market share throughout the product life-cycle.
  
\begin{figure}[htbp]
\centering
   \begin{subfigure}{}
    \includegraphics[width=5cm]{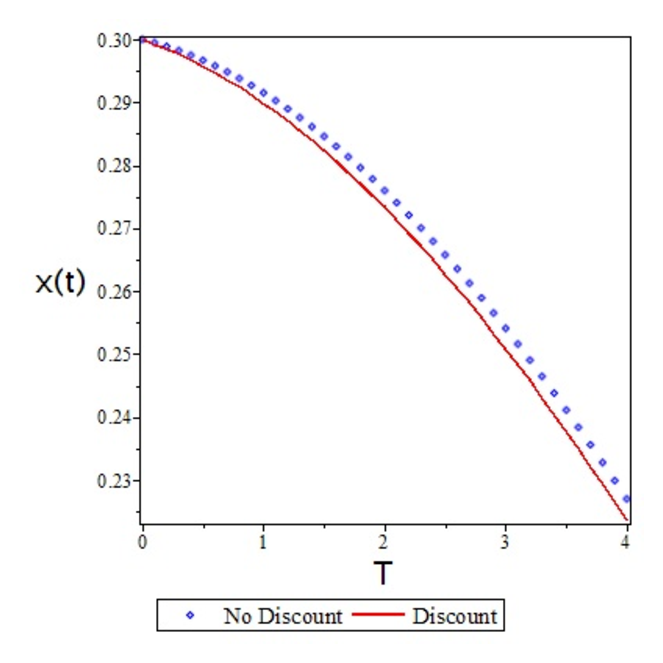}
   \end{subfigure}
\vspace{.5cm}
   \begin {subfigure}{}
\centering
     \includegraphics[width=5cm]{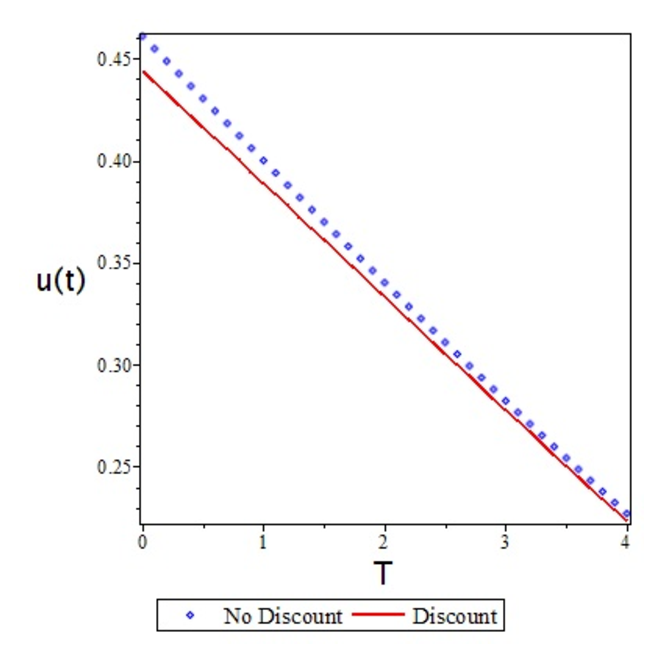}
   \end{subfigure}
\caption{Market conditions with $x_0 = 0.3$}\label{Fig:base03}

\end{figure}

In Figure \ref{Fig:base03}, we see the optimal control path for maintenance and market share for the firm under Case 2 of Corollary \ref{cor:3cases}.  When $x_0 = 0.3$, it is no longer profitable to rigorously maintain the product; as a result, market share monotonically decreases for the product life-cycle.  Case 2 represents a vendor with small market share who wishes to maximize revenue without incurring high costs.  In our model, it is too costly for this firm to gain market share.  Instead, the firm capitalizes on short-term profits before market share deteriorates.  Maintenance effort is much lower at every point in time in Case 2 compared to Case 1.

\begin{figure}[htbp]
\centering
   \begin{subfigure}{}
     \includegraphics[width=5cm]{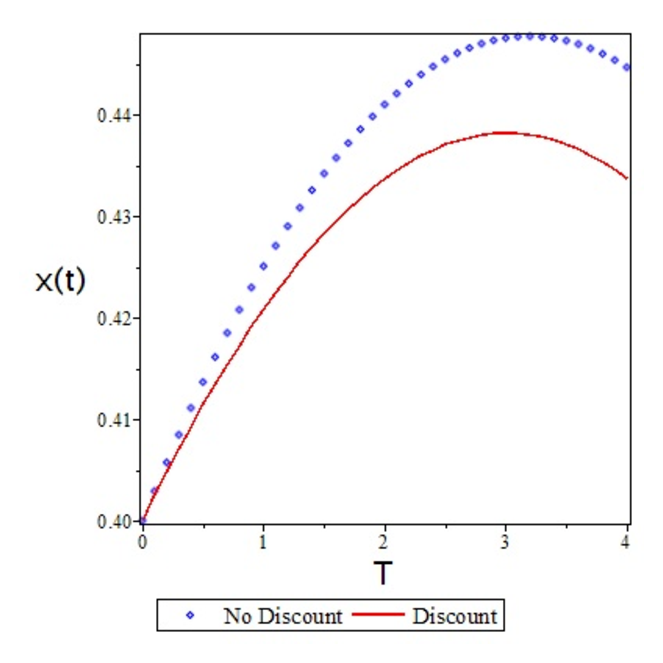}
   \end{subfigure}
\vspace{.5cm}
   \begin {subfigure}{}
\centering
     \includegraphics[width=5cm]{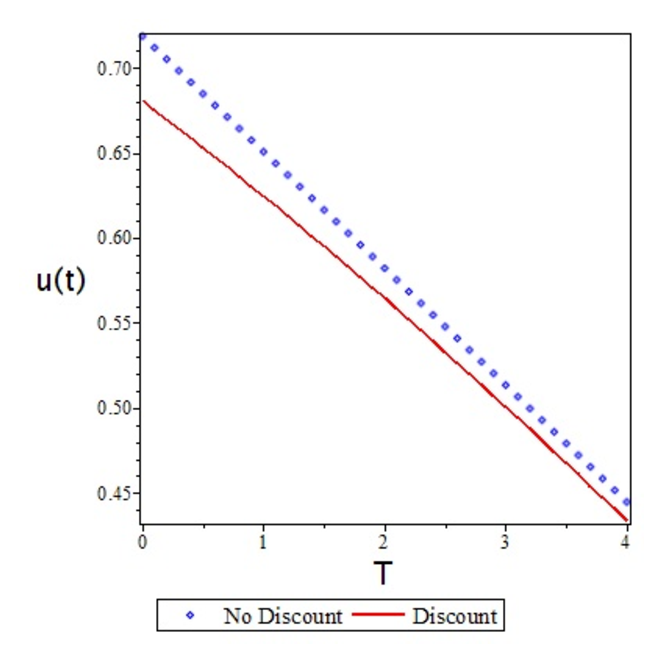}
   \end{subfigure}
\caption{Market conditions with $x_0 = 0.4$}\label{Fig:base04}
\end{figure}

In Figure \ref{Fig:base04}, we see the optimal control path for product maintenance and market share for the firm under Case 3 of Corollary \ref{cor:3cases}.  Because numerical examples are sensitive to initial values, a decrease in $x_0$ of $10\%$ is sufficient to shift both the optimal control path and market share behavior.  There is less focus on maintenance at every time $t$. Market share first increases, peaks, then decreases.  From Corollary \ref{cor:3cases}, we know that the peak is reached when $u = \frac{\pi_0}{P}$.  Unlike the first case, the vendor is not incentivized to sustain the necessary maintenance effort for constant market growth.  In this case, focus on product maintenance lies between efforts in Case 1 and 2.

\section{Conclusion and Future Directions}
In this paper, we propose a parsimonious optimal control problem to model a firm's dynamic decision-making process during software development and distribution.  We examine the relationship between market share and product maintenance when operating in a digital market. Our theoretical analysis demonstrates that market share behavior can be characterized as a function of product maintenance dedication and firm specific conditions; though market share may increase or decrease over the product life-cycle, maintenance effort should always be decreasing as long as the perceived value of the product is above a certain threshold.  This finding is independent of initial conditions and the salvage value of end-of time market share, providing an analytic foundation for Lehman's 7th law of software evolution. 

To help visualize our findings, we employ numerical examples of the linear case when maintenance effort is always decreasing over time. These numerical illustrations emphasize the different optimal control paths under different market conditions.  As expected, initial market share and cost of product maintenance play critical roles in the vendor's decision-making process.  Our theoretical results are robust to discounting as well an innovation coefficient.

For future work, it would be instructive to determine whether the sufficient conditions for optimality can be simplified even further or if this problem is inherently difficult to confirm the optimality of a solution to the derived necessary conditions. Additional model extensions would include incorporating the market share into the cost of production $C$ or considering a more general revenue function $R(u,x)$. It would also be interesting to consider the problem of market share gain in an even more fundamental epidemic model.

We also look to extend our research by analyzing the case with multiple vendors in a digital market setting.  The addition of other firms adds competition, changing the dynamics of the model from an optimal control problem to a differential game, as each firm must account for the actions of all other firms at every time $t$.  We hope to better understand how our results hold up in this game-theoretic setting. Another avenue of research involves parameterizing peer review feedback as an endogenous variable that firms can influence through marketing and focus on product maintenance.  This would add to the richness of the model and may help better fit empirical data, at the cost of a more concise model.

\bibliography{library}

\begin{thebibliography}{10}
\expandafter\ifx\csname url\endcsname\relax
  \def\url#1{\texttt{#1}}\fi
\expandafter\ifx\csname urlprefix\endcsname\relax\def\urlprefix{URL }\fi
\expandafter\ifx\csname href\endcsname\relax
  \def\href#1#2{#2} \def\path#1{#1}\fi

\bibitem{Apple}
{Apple Press Info},
  \href{https://www.apple.com/pr/library/2014/01/27Apple-Reports-First-Quarter-Results.html}{{Apple
  - Press Info - Apple Reports First Quarter Results}} (2014).
\newline\urlprefix\url{https://www.apple.com/pr/library/2014/01/27Apple-Reports-First-Quarter-Results.html}

\bibitem{Lehman1997}
M.~Lehman, J.~Ramil, P.~Wernick, D.~Perry, W.~Turski, {Metrics and laws of
  software evolution-the nineties view}, Proceedings Fourth International
  Software Metrics Symposium~(4th International Software Metrics Symposium).
\newblock \href {http://dx.doi.org/10.1109/METRIC.1997.637156}
  {\path{doi:10.1109/METRIC.1997.637156}}.

\bibitem{Wolverton1974}
R.~W. Wolverton, {The Cost of Developing Large-Scale Software}, Computers, IEEE
  Transactions on C-23~(6) (1974) 615--636.
\newblock \href {http://dx.doi.org/10.1109/T-C.1974.224002}
  {\path{doi:10.1109/T-C.1974.224002}}.

\bibitem{Zahedi1991}
F.~Zahedi, N.~Ashrafi, {Software reliability allocation based on structure,
  utility, price, and cost}, IEEE Transactions on Software Engineering 17~(4)
  (1991) 345--356.
\newblock \href {http://dx.doi.org/10.1109/32.90434}
  {\path{doi:10.1109/32.90434}}.

\bibitem{Munson1995}
J.~C. Munson, {Software measurement: Problems and practice}, Annals of Software
  Engineering 1~(1) (1995) 255--285.
\newblock \href {http://dx.doi.org/10.1007/BF02249053}
  {\path{doi:10.1007/BF02249053}}.

\bibitem{Johari2011}
K.~Johari, A.~Kaur, {Effect of software evolution on software metrics}, in: ACM
  SIGSOFT Software Engineering Notes, Vol.~36, 2011, p.~1.
\newblock \href {http://dx.doi.org/10.1145/2020976.2020987}
  {\path{doi:10.1145/2020976.2020987}}.

\bibitem{Drouin2013}
N.~Drouin, M.~Badri, {Investigating the Applicability of the Laws of Software
  Evolution : A Metrics Based Study}, in: Evaluation of Novel Approaches to
  Software Engineering, Springer Berlin Heidelberg, 2013, pp. 174--189.

\bibitem{Neamtiu2012}
I.~Neamtiu, G.~Xie, J.~Chen, {Towards a better understanding of software
  evolution: an empirical study on open-source software}, Journal of
  Software-Evolution and Process 24~(September 2011) (2012) 481--491.
\newblock \href {http://dx.doi.org/10.1002/smr} {\path{doi:10.1002/smr}}.

\bibitem{Yu2013}
L.~Yu, A.~Mishra, {An Empirical Study of Lehman's Law on Software Quality
  Evolution}, International Journal of Software Informatics 7~(3) (2013)
  469--481.

\bibitem{Zhang2013}
J.~Zhang, S.~Sagar, E.~Shihab, {The evolution of mobile apps: an exploratory
  study}, in: DeMobile' 13, 2013, pp. 1--8.
\newblock \href {http://dx.doi.org/10.1145/2501553.2501554}
  {\path{doi:10.1145/2501553.2501554}}.

\bibitem{Ji2005}
Y.~Ji, V.~S. Mookerjee, S.~P. Sethi,
  \href{http://pubsonline.informs.org/doi/abs/10.1287/isre.1050.0059}{{Optimal
  Software Development: A Control Theoretic Approach}}, Information Systems
  Research 16~(3) (2005) 292--306.
\newblock \href {http://dx.doi.org/10.1287/isre.1050.0059}
  {\path{doi:10.1287/isre.1050.0059}}.
\newline\urlprefix\url{http://pubsonline.informs.org/doi/abs/10.1287/isre.1050.0059}

\bibitem{Ji2011}
Y.~Ji, S.~Kumar, V.~S. Mookerjee, S.~P. Sethi, D.~Yeh, {Optimal enhancement and
  lifetime of software systems: A control theoretic analysis}, Production and
  Operations Management 20~(6) (2011) 889--904.
\newblock \href {http://dx.doi.org/10.1111/j.1937-5956.2010.01215.x}
  {\path{doi:10.1111/j.1937-5956.2010.01215.x}}.

\bibitem{Haruvy2008}
E.~Haruvy, S.~P. Sethi, J.~Zhou, {Open source development with a commercial
  complementary product or service}, Production and Operations Management
  17~(1) (2008) 29--43.
\newblock \href {http://dx.doi.org/10.3401/poms.1070.0004}
  {\path{doi:10.3401/poms.1070.0004}}.

\bibitem{He2008}
X.~He, A.~Prasad, S.~P. Sethi, {Cooperative advertising and pricing in a
  dynamic stochastic supply chain: Feedback stackelberg strategies}, PICMET:
  Portland International Center for Management of Engineering and Technology,
  Proceedings 18~(1) (2008) 1634--1649.
\newblock \href {http://dx.doi.org/10.1109/PICMET.2008.4599783}
  {\path{doi:10.1109/PICMET.2008.4599783}}.

\bibitem{He2011}
X.~He, A.~Krishnamoorthy, A.~Prasad, S.~P. Sethi,
  \href{http://dx.doi.org/10.1016/j.orl.2010.10.006}{{Retail competition and
  cooperative advertising}}, Operations Research Letters 39~(1) (2011) 11--16.
\newblock \href {http://dx.doi.org/10.1016/j.orl.2010.10.006}
  {\path{doi:10.1016/j.orl.2010.10.006}}.
\newline\urlprefix\url{http://dx.doi.org/10.1016/j.orl.2010.10.006}

\bibitem{B78}
J.~Baillieul, Geometric methods for nonlinear optimal control problems, J.
  Optimization Theory and Applications 25~(4) (1978) 519--548.

\bibitem{Lahiri2013}
A.~Lahiri, D.~Dey, {Effects of Piracy on Quality of Information Goods},
  Management Science~(July 2014).
\newblock \href {http://dx.doi.org/10.1287/mnsc.1120.1578}
  {\path{doi:10.1287/mnsc.1120.1578}}.

\bibitem{Bass2}
F.~M. Bass, T.~V. Krishnan, D.~C. Jain, Diffusion of new products: Empirical
  generalizations and managerial uses, Marketing Science 14~(3) (1995)
  G79--G88.

\bibitem{Bass1994}
F.~M. Bass, T.~V. Krishnan, D.~C. Jain, {Why the Bass Model Fits without
  Decision Variables}, Marketing Science 13~(3) (1994) 203--223.
\newblock \href {http://dx.doi.org/10.1287/mksc.13.3.203}
  {\path{doi:10.1287/mksc.13.3.203}}.

\bibitem{Bass2004}
F.~M. Bass, {Comments on A New Product Growth for Model Consumer Durables},
  Management Science 50 (2004) 1833--1840.
\newblock \href {http://dx.doi.org/10.1287/mnsc.1040.0300}
  {\path{doi:10.1287/mnsc.1040.0300}}.

\bibitem{Bass2004a}
F.~M. Bass, {A New Product Growth for Model Consumer Durables}, Management
  Science 50~(12) (2004) 1825--1832.
\newblock \href {http://dx.doi.org/10.1287/mnsc.1040.0264}
  {\path{doi:10.1287/mnsc.1040.0264}}.

\bibitem{J70}
D.~H. Jacobson, Sufficient conditions for non-negativity of the second
  variation in singular and non-singular control problems, SIAM J. Control
  8~(3).

\bibitem{BH65}
J.~V. Breakwell, Y.-C. Ho, On the conjugate point condition for the control
  problem, Int. J. Engng. Sci. 2 (1965) 565--579.

\bibitem{Mangasarian1966}
O.~L. Mangasarian, {Sufficient Conditions for the Optimal Control of Nonlinear
  Systems}, SIAM Journal on Control 4~(1) (1966) 139--152.
\newblock \href {http://dx.doi.org/10.1137/0304013}
  {\path{doi:10.1137/0304013}}.

\bibitem{P71}
D.~W. Peterson, A sufficient maximum principle, IEEE Trans. Automatic Control
  (1971) 85--86.

\bibitem{KS71}
M.~I. Kamien, N.~L. Schwartz, {Sufficient Conditions in Optimal Control
  Theory}, J. Economic Theory 3 (1971) 207--214.

\bibitem{SS77}
A.~Seierstad, K.~Sydsaeter,
  \href{http://www.jstor.org/stable/2525753}{Sufficient conditions in optimal
  control theory}, International Economic Review 18~(2) (1977) 367--391.
\newline\urlprefix\url{http://www.jstor.org/stable/2525753}

\bibitem{Z84}
V.~Zeidan, First and second order sufficient conditions for optimal control and
  the calculus of variations, Appl. Math. Optim. 11 (1984) 209--226.

\bibitem{MO02}
H.~Maurer, H.~J. Oberle, {Second order sufficient conditions for optimal
  control problems with free final time: the Riccati approach}, SIAM J. Control
  Optim. 41~(2) (2002) 380--403.

\end{thebibliography}
\bibliographystyle{elsarticle-num}

\end{document}